\documentclass[twocolumn,showpacs,preprintnumbers,amsmath,amssymb]{revtex4}
\usepackage{graphicx}% Include figure files
\usepackage{dcolumn}% Align table columns on decimal point
\usepackage{bm}% bold math

\newcommand{\be}{\begin{eqnarray}}
\newcommand{\ee}{\end{eqnarray}}

\begin{document}

\preprint{APS/123-QED}

\title{Finite-temperature phase structures of hard-core bosons in an optical lattice with an effective magnetic field}
% Force line breaks with \\

\author{Yuki Nakano, Kenichi Kasamatsu, and Tetsuo Matsui}
% \altaffiliation[Also at ]{}%Lines break automatically or can be forced with \\
%\author{${}^1$Kenichi Kasamatsu${}^2$}%
%\author{Makoto Tsubota${}^1$}%
% \email{hiromitu@sci.osaka-cu.ac.jp}
\affiliation{%
Department of Physics, Kinki University, Higashi-Osaka, Osaka 577-8502, Japan
}%
%\affiliation{
\date{\today}% It is always \today, today,

\begin{abstract}
We study finite-temperature phase structures of hard-core bosons in a two-dimensional optical lattice
subject to an effective magnetic field by employing the gauged CP$^1$ model.
Based on the extensive Monte Carlo simulations, we study their phase 
structures at finite temperatures for several values of the magnetic flux 
per plaquette of the lattice and mean particle density. 
Despite the presence of the particle number fluctuation, the thermodynamic 
properties are qualitatively similar to those of the frustrated XY model with 
only the phase as a dynamical variable. This suggests that 
cold atom simulators of the frustrated XY model are available irrespective 
of the particle filling at each site. 
\end{abstract}

\maketitle
\section{Introduction} \label{intro}
Ultracold atoms in an optical lattice (OL) have been a particularly important field to 
study a wide range of fundamental problems in condensed matter physics \cite{Bloch}. 
When an OL is rotated,  we can expect versatile cold-atom quantum simulators, 
which can demonstrate various effects caused by a magnetic field 
such as quantum Hall effects \cite{Cooper}. 
This is because the neutral atoms in a rotating reference frame experience 
a Coriolis force of the same form as the Lorentz
force on charged particles in a magnetic field.
Recently, two experiments were reported, making use of a rotating OL 
to study quantized vortices in gaseous Bose-Einstein condensates 
(BECs) \cite{Tung,Williams}. 
Moreover, ``synthesis" of a gauge field was realized by 
using a spatially varying Raman coupling between internal atomic states 
to implement
the required geometric phases \cite{Dalibard,Lin,Lin2,Aidelsburger}, 
opening a door to 
more wide ranging studies associated with an artificial magnetic field 
that gives rise to an orbital motion of neutral atoms. 

The Bose-Hubbard model under an effective magnetic field exhibits 
very interesting physics \cite{Jaksch,Mueller,Sorensen,Gerbier,Oktel,Umucalilar1,Goldbaum,
Palmer,Polak,Moller,Umucalilar2,Duric,Powell,Moller2,Zhang}, 
far beyond the physics of the usual Bose-Hubbard model. 
These properties are inherited from the remarkable
structure of the energy spectrum for noninteracting problems,
i.e., a single particle moving on a tight-binding lattice in the
presence of a uniform magnetic field. Here, 
the energy spectrum depends sensitively on the frustration parameter $f$, 
a magnetic flux per plaquette 
of the lattice, and exhibits a fractal structure known as the 
``Hofstadter butterfly" \cite{Hofstadter}. 
For a rational value $f=p/q$ (with the integers $p$ and $q$), there are 
$q$ bands, and each state is $q$-fold degenerate \cite{Brown,Zak}. 
%Recent works have studied the detailed properties of the 
%ground state and its low-energy excitations \cite{Duric,Powell,Moller2} and the superfluid-Mott transition 
%\cite{Oktel,Umucalilar1,Goldbaum,Polak,Umucalilar2,Sachdeva} of this model. 
A series of theoretical proposals indicates that 
it should be possible to implement
strong gauge fields such as $f \sim 1$ on 
an OL \cite{Jaksch,Mueller,Sorensen,Gerbier}. In fact, it has been
experimentally demonstrated recently \cite{Aidelsburger}. 
In the strongly interacting regime, where both the particle number 
per site and the magnetic flux per plaquette are of order unity, 
it is theoretically predicted 
that there exist strongly correlated phases representative
of the continuum quantum Hall states \cite{Sorensen,Palmer,Moller}. 
In the weakly-interacting condensed phase under the strong magnetic field, 
the frustrated Josephson junction arrays provide a close analog of that system, and 
implementations using cold atoms have been proposed \cite{Polini,Kasamatsu2}. 

These studies focused mainly on the ground-state properties 
and the quantum phase transitions at {\it zero temperature}.
However, they are mostly based on 
the exact diagonalization study and are restricted to small systems.
In this paper, we study the {\it finite-temperature} phase diagram of 
hard-core bosons 
in a two-dimensional (2D) OL with an effective gauge field. 
Our study is relevant to atoms with low densities and very strong repulsive 
interactions, e.g., tuned by the Feshbach resonance \cite{Chin}. 
In the hard-core limit, the Bose-Hubbard model can be mapped to the quantum 
spin model and described by the CP$^1$ (complex projective) 
operators, which are useful to construct the path-integral formulation \cite{Nakano}. 
In the high-temperature limit, the quantum Hamiltonian reduces to the 
classical {\it three-component} XY model [i.e., without nearest-neighbor 
$s_z$ coupling as shown in Eq.(\ref{eq:spham}) below] frustrated by the gauge field, which is 
referred to as the gauged CP$^1$ 
model below. This reduction provides a practical platform 
to explore the finite-temperature phase diagram of this system and to discuss the detailed 
critical properties of the phase transitions. 

It is known that the 2D frustrated XY model (FXYM), which has
only two components $s_x$ and $s_y$, is closely related 
to our model and  exhibits very rich properties of the phase transition 
\cite{TJ83,LKG,Santiago-Jose,LeeLee,Olsson,Luo,Korshunov,HPV05a,Olsso1n-Teitel,Minnhagen,Okumura}. 
For the fully frustrated XY model (FFXYM) with $f=1/2$, there would be double phase transitions, one is  associated with 
the Berezinskii-Kosterlitz-Thouless (BKT) transition 
due to the global U(1) symmetry 
and the other associated with Ising-model-like 
transition due to the Z$_2$ chiral symmetry of the ground state.  
A central issue in the studies of the FFXYM has been to clarify how these 
two distinct types of orderings take place \cite{TJ83}. 
One possibility is that, even at the temperature 
where the  Z$_2$ chirality establishes a long-range order at $T < T_c$, 
the U(1) phase (XY-spin) may 
remain disordered due to thermally excited, unbound vortices. 
Then, the orderings of the two variables take place at 
two separate temperatures such that $T_{c} > T_{\rm BKT}$. 
The other possibility is that both orderings of the chirality and 
the phase take place at the same temperature, and 
the resulting single phase transition is neither of   
the conventional Ising type nor the BKT type 
but follows a new universality class.
Many studies have discussed this phase transition, 
but yet do not provide conclusive results. 
The phase transitions for other values of $f$, e.g., $f=1/3$, 2/5, etc., 
were also discussed 
in Refs. \cite{Grest,Li,Lee,Denniston}. 

Our study is an extension of these studies by incorporating the particle number 
fluctuation ($z$-component of the spin) at each site. It is nontrivial how the additional 
degree of freedom has an influence on the above thermodynamic properties. 
To the best of our knowledge, the statistical properties of this model have not been 
considered so far. 
We confine ourselves to the values of the frustration parameter $f=0,$ 1/2, and 2/5 
to settle the discussion, and make a comparison with the 
results obtained by the usual XY model. Although our concerning model is 
a classical one (obtained by neglecting the quantum fluctuations relevant at 
low temperatures), 
this should properly describe the thermodynamic properties of the hard-core 
bosons in an optical lattice at sufficiently high temperatures \cite{Shimizu}.  

In Sec.\ref{formulation}, we describe how to obtain our gauged CP$^1$ model, starting from 
the Bose-Hubbard Hamiltonian with the gauge field. The ground-state properties of 
the gauged CP$^1$ model are discussed in Sec. \ref{ground}. In Sec. \ref{phasestructure}, based on 
the Monte-Carlo simulations, we study the finite-temperature phase structures 
of the gauged CP$^1$ model for the averaged site occupation 
of hard-core bosons being half-filled and non-half-filled.  
Section \ref{conclusion} is devoted to conclusions and discussion. 

\section{Model} \label{formulation}
\subsection{Hamiltonian for hard-core bosons in an effective magnetic field}
We consider a system of $N$-bosons put on the sites of a 
2D square lattice with the size $L\times L$.
The 2D Bose-Hubbard model
subject to a uniform Abelian gauge potential is described by the Hamiltonian
\begin{eqnarray}
\nonumber
\hat{H} = - \frac{t}{2} \sum_{\langle i,j\rangle } \left[\hat{a}_i^\dag\hat{a}_j e^{i A_{ij}} +
  {\rm h.c.} \right] + \frac{U}{2}\sum_i \hat{\rho}_i(\hat{\rho}_i-1) \\
- \mu\sum_i \hat{\rho}_i.
\label{eq:hamiltonian}
\end{eqnarray}
Here, the operator $\hat{a}_i^{(\dag)}$ destroys (creates) a boson on the lattice
site $i = (i_x,i_y)$. $\langle i,j\rangle $ implies a 
pair of nearest-neighbor sites. $t$, $\mu$, and $U( \geq 0)$ describe 
the nearest-neighbor tunneling energy, the chemical potential, and the
onsite repulsion, respectively. 
$\hat{\rho}_i=\hat{a}^\dag_i\hat{a}_i$ is the
number operator at $i$
and the Hamiltonian conserves the total number of bosons,
$\hat{N} = \sum_i \hat{\rho}_i$. 
Throughout this work, we consider the system without additional trap 
potentials such as a harmonic one. Nevertheless, the results of this study 
can be applied within the local density approximation to realistic
experimental systems which have an additional trapping potential.

The field $A_{ij}$ describes the imposed
gauge potential, defined by $A_{ij} = \int_{{\bf r}_{j}}^{{\bf r}_{i}} {\bf A} \cdot d {\bf r}$. 
All of the physics of the system governed by the Hamiltonian
(\ref{eq:hamiltonian}) is gauge-invariant. Hence, its properties depend 
only on the magnetic fluxes of magnitude $B$ through plaquettes
\begin{equation} 
\Phi = \int_{\rm plaq} d {\bf S} \cdot {\bf B} 
= \sum_{i,j\in \alpha} A_{ij} = B d^2 =2 \pi f,
\label{eq:flux}
\end{equation}
where $\alpha$ labels the plaquette, and the sum represents the
directed sum of the gauge fields around that plaquette (the discrete
version of the line integral). 
$d$ is a typical lattice spacing and the last equality relates
$B$ and $f$. In the following, 
we use the vector potential in the symmetric gauge and its form can be
written as 
\begin{equation}
{\bf A}=(A_x,A_y,A_z) = \left( -\frac{B}{2}y, \frac{B}{2}x, 0 \right). 
\end{equation}
This corresponds to a uniform magnetic field 
${\bf B}= (0,0, B)$ in the direction perpendicular to the lattice plane.
The discrete vector potential is then written as 
\begin{eqnarray}
A_{ij} = \biggl\{
\begin{array}{c}
  - \pi f i_{y}  \hspace{3mm} {\rm for} \hspace{2mm} i=(i_x,i_y), \hspace{1mm} j=(i_{x}+1,i_y)  \\
   \pi f i_{x}  \hspace{3mm} {\rm for} \hspace{2mm} i=(i_x,i_y), \hspace{1mm} j=(i_{x},i_{y}+1). 
\end{array}
\end{eqnarray} 

In the following, we take a hard-core limit ($U \rightarrow \infty$), 
where the allowed physical states at $i$ are eigenstates
of $\hat{\rho}_i$ with eigenvalue 0 or 1 and their superpositions.
States  with higher particle number
at the same site such as double occupancy are excluded. We introduce 
a destruction operator of the hard-core boson as $\hat{\phi}_i$, which 
satisfies the following mixed canonical-(anti)commutation relations 
\begin{eqnarray}
\left[\hat{\phi}_i, \hat{\phi}_j \right] = 0 , \hspace{3mm}
\left[\hat{\phi}_i, \hat{\phi}^{\dagger}_j \right]  = 0 \ \ {\rm for} \ \ i \neq j, 
\end{eqnarray}
and on the same site, 
\begin{eqnarray}
\left\{ \hat{\phi}_i, \hat{\phi}_i \right\} = 0,\hspace{3mm}
\left\{ \hat{\phi}_i, \hat{\phi}^\dagger_i \right\} = 1.
\end{eqnarray}
Thus,  the number 
operator is rewritten as $\hat{\rho}_{i} = \hat{\phi}^\dag_i \hat{\phi}_i$
and its eigenvalue $\rho_i$ is  assured to be  
0 or 1.
%and the eigenstates are denoted as $| \rho_i  = 0 \rangle$ 
%and $| \rho_i  = 1 \rangle$. 
Then, the Hamiltonian $\hat{H}$ is written as
\begin{equation}
\hat{H}_{\rm hc} = -\frac{t}{2} \sum_{\langle i,j \rangle} \left(\hat{\phi}_{i}^\dag \hat{\phi}_j e^{iA_{ij}}
+{\rm H.c.}\right)-\mu \sum_i \hat{\phi}_i^\dag \hat{\phi}_i,  \label{hardcorehamil}
\end{equation}
where $\mu$ determines the mean density $\rho$ of hard-core bosons per site,
\begin{equation}
\rho\equiv  \langle \bar{\rho} \rangle, \hspace{5mm}  \bar{\rho} \equiv 
\frac{1}{L^2} \sum_i \hat{\phi}_i^\dag \hat{\phi}_i, 
\label{meandensity}
\end{equation}
within the range $0\leq \rho \leq 1$.
%where $=\sum_i \hat{\rho}_{i} / L^2$ represents 
%the site average of the density.

In the hard-core limit, the Bose-Hubbard model becomes equivalent to 
a spin-1/2 quantum magnet, because the following relations exist 
between $\hat{\phi}_i$ and the $s=1/2$ SU(2) spin operator $\hat{s}_i^{x,y,z}$ \cite{Nakano},
\begin{eqnarray}
&&\hat{s}_i^z = \hat{\phi}_i^{\dag} \hat{\phi}_i-\frac{1}{2}, \hspace{5mm} 
\hat{s}_i^+ \equiv \hat{s}_i^x+i\hat{s}_i^y= \hat{\phi}_i^\dag, \hspace{5mm}
\hat{s}_i^- = \hat{\phi}_i,\nonumber\\
&&(\hat{s}_i^x)^2+(\hat{s}_i^y)^2+(\hat{s}_i^z)^2 =\frac{3}{4}.
 \label{spinphirelation}
\end{eqnarray}
The Hamiltonian (\ref{hardcorehamil}) then becomes
\begin{eqnarray}
\hspace{-0.6cm}
\hat{H}_{\rm hc} =  -\frac{t}{2} \sum_{\langle i,j\rangle} \left( \hat{s}_i^+\hat{s}^-_j e^{i A_{ij}} +
   \hat{s}_j^+\hat{s}^-_i e^{-i A_{ij}} \right)  - \mu\sum_i \hat{s}^z_i.
\label{eq:spham}
\end{eqnarray}
Here, the conservation of total particle number is interpreted 
as the constant magnetization $\hat{S}^z = \sum_i \hat{s}^z_i$. 
Under Eq.(\ref{spinphirelation}), the eigenstates have 
the correspondence $|\!\!\uparrow_{i} \rangle = |\rho_{i}= 1 \rangle $ 
and $|\!\!\downarrow_{i} \rangle = |\rho_{i}= 0 \rangle $.
This Hamiltonian describes a quantum 
spin-1/2 magnet, experiencing XY nearest neighbor spin exchange interactions. 
These exchange interactions are frustrated due to the gauge field $A_{ij}$.  

\subsection{CP$^1$ variable and path-integral representation}
To study the thermodynamic properties of this system, 
we evaluate the partition function $Z$ of the grand canonical ensemble,
\begin{equation} 
Z = {\rm Tr} e^{-\beta \hat{H}_{\rm hc}}, \hspace{5mm} \beta = \frac{1}{k_B T}.
\end{equation}
We express $Z$ by a path integral which is useful for numerical calculations.
For this purpose, it is convenient to introduce a CP$^{1}$ variable 
$w_{i} = (w_{1i}, w_{2i}) \in C$ which 
satisfies the CP$^1$ constraint,
\begin{equation}
|w_{1i}|^2+|w_{2i}|^2=1.
\label{eq:cp1}
\end{equation}
An associated pseudocoherent state $| w_i \rangle$ is defined by
\begin{equation}
| w_i \rangle \equiv w_{1i} | \uparrow_i \rangle + w_{2i} | \downarrow_i \rangle,
\end{equation}
where $| w_i \rangle$ is normalized as $\langle {w}_{i} | w_{i} \rangle =1$ due to 
Eq. (\ref{eq:cp1}), and  we have generally  
$\langle {w}_{i} | w_{i}' \rangle = w_{i1}^{\ast} w_{i1} '+ w_{i2}^{\ast} 
w_{i2}' = w_i^\ast w_i'$. 
Let us define the integration measure 
\begin{equation}
\int [d^2w_{i}] \equiv 2 \int_C d^2 w_{1i} \int_C d^2 w_{2i} \delta \left( \langle {w}_{i} | w_{i} \rangle -1 \right) 
\end{equation}
which satisfies 
\begin{equation}
\int [d^2w_{i}] 1=2,\hspace{3mm} \int [d^2w_{i}]w_{ia}^\ast w_{ib}=
2\times\frac{1}{2}\delta_{ab}=\delta_{ab}. 
\end{equation}
Then the completeness is expressed as
\begin{equation}
\int [d^2w_{i}] | w_{i} \rangle \langle {w}_{i} | =  
| \uparrow _{i} \rangle \langle \uparrow _{i} | 
+  | \downarrow_{i} \rangle \langle \downarrow_{i} | = 1.
\end{equation}

The Hamiltonian can be represented by the CP$^1$ operators $\hat{w}_{1i}$ 
and $\hat{w}_{2i}$ which satisfy the bosonic commutation relation 
\begin{equation}
[ \hat{w}_{ai} , \hat{w}_{bj}] = 0, \hspace{3mm} [ \hat{w}_{ai} , 
\hat{w}^{\dag}_{bj}] = \delta_{ab} \delta_{ij},  \hspace{3mm} 
a,b =1,2,
%(a,b) \in (1,2),
\end{equation}
and their physical states are restricted as
$\sum_a\hat{w}^\dag_{ai} \hat{w}_{ai}|{\rm phys}\rangle =
|{\rm phys}\rangle$.
Here, the correspondence to the spin operator is given as 
\begin{equation}
|\!\uparrow_{i} \rangle = \hat{w}_{1i}^{\dag} | {\rm vac} \rangle, \hspace{5mm} |\!\downarrow_{i} \rangle = \hat{w}_{2i}^{\dag} | {\rm vac} \rangle,
\end{equation}
and
\begin{eqnarray}
\hat{s}^{x,y,z}_{i}=\frac{1}{2}(\hat{w}_{1i}^{\dag},\hat{w}_{2i}^{\dag})
\sigma^{x,y,z}(\hat{w}_{1i},\hat{w}_{2i})^t,\nonumber\\
\hat{s}^{z}_{i} = \hat{w}_{1i}^{\dag} \hat{w}_{1i} - \frac{1}{2}, \hspace{3mm} 
\hat{s}^{+}_{i} = \hat{w}_{1i}^{\dag} \hat{w}_{2i}, \hspace{3mm} 
\hat{s}^{-}_{i} = \hat{w}_{2i}^{\dag} \hat{w}_{1i},\label{spinwrelation}
\end{eqnarray}
where $\sigma^{x,y,z}$ are Pauli matrices.
Equations (\ref{spinwrelation}) and (\ref{spinphirelation}) imply the relation
$\hat{\phi}_i=\hat{w}_{2i}^{\dag} \hat{w}_{1i}$, etc.
We also note the following relation :
\begin{equation}
\langle w_i'|\hat{w}_{ia}^\dag
\hat{w}_{ib}| w_i \rangle=(w_{ia}')^{\ast} w_{ib}.
\end{equation}

Using these relations and following the standard procedure,
we can write the partition function $Z$ 
in the path-integral form with the imaginary time $\tau \in [0,\beta]$ as 
\begin{equation} 
Z =  \prod_{i,\tau} \int [d^2 w_{i}(\tau)] e^{ \int_0^\beta d\tau A(\tau)}, 
\end{equation}
where 
\begin{eqnarray}
A(\tau) =  - \sum_{i,a} {w}_{ai}^{\ast}(\tau)\dot{w}_{ai}(\tau) \nonumber \\
+  \frac{t}{2} \sum_{\langle i,j \rangle} \left( w_{1i}^{\ast}(\tau) w_{2i}(\tau) 
w_{2j}^{\ast}(\tau) w_{1j}(\tau) e^{i A_{ij}} + {\rm c.c.} \right) \nonumber \\
+ \mu \sum_{i} w_{1i}^{\ast}(\tau) w_{1i}(\tau) .
\label{2+1d}
\end{eqnarray}
Here, the CP$^1$ operators have been replaced to the complex numbers 
by employing the path-integral formulation. 

To proceed further, we make one simplification by considering the finite-$T$ region, 
such that the $\tau$-dependence of $w_{ai}(\tau)$ in the path integral can 
be ignored, keeping only the zero modes as $w_{ai}(\tau) \to w_{ai}$ \cite{Shimizu}. 
This corresponds to neglecting the quantum fluctuations. 
%and ignore the $\tau$-dependence of $w_{ai}(\tau)$.
%Then, $Z$ is simply evaluated only from the contribution of the zero mode of $w(\tau)$. 
As a result, the problem is reduced to the classical one. 
Under the relations (\ref{spinphirelation}) and (\ref{spinwrelation}), one can 
finally obtain
\begin{equation}
Z= \prod_{i} \int [d^2 w_i] e^{-\beta H_{\rm CP^{1}}(w)},  \label{2d}  
\end{equation}
where $H_{\rm CP^{1}}(w)$ is the classical version 
of Eq. (\ref{hardcorehamil}) expressed in terms of $w_i$: 
\begin{eqnarray}
H_{\rm CP^{1}}(w) &=&-\frac{t}{2} \sum_{\langle i,j \rangle} \left( 
\phi_{i}^{\ast} \phi_{j} e^{i A_{ij}} + {\rm c.c.} \right) - \mu \sum_{i} 
w_{1i}^{\ast} w_{1i},\nonumber\\
\phi_i&\equiv&w_{2i}^\ast w_{1i}.
\label{2denergy}
\end{eqnarray}
This allows us to study the finite-$T$ phase structure, which 
summarizes the essential properties of the system. Besides, the finite-$T$ 
phase diagram gives a very useful insight into the phase structure at $T=0$; 
if some ordered states are found at finite $T$, we can naturally expect 
that they persist down to $T=0$. 
Equation (\ref{2denergy}) is referred to as the ``gauged CP$^1$ model" in the following \cite{gauge}. 
We can evaluate the partition function of Eq. (\ref{2d}) to understand the thermodynamic properties of 
the system, using the standard Monte Carlo simulations. 
From the symmetry properties summarized in Appendix \ref{app1}, 
we can confine ourselves to $0 \leq f \leq 1/2$ and $1/2 \leq \rho \leq 1$ in the 
following argument. 

We note that Eq.(\ref{2denergy}) is viewed as a gauged model of three-component
normalized classical spin $\vec{s}_i$ [O(3) spin] \cite{O3} as 
\begin{eqnarray}
\hspace{-0.4cm}
\vec{s}_i&\equiv& w_{i}^\dag\vec{\sigma} w_i,\quad
\vec{s}_i \cdot \vec{s}_i=1,\nonumber\\
\hspace{-0.4cm}
H_{\rm CP^{1}}(w)&=&
-\frac{t}{8} \sum_{\langle i,j \rangle} \left( 
s_{i}^+ s_{j}^- e^{i A_{ij}} + {\rm c.c.} \right) - \frac{\mu}{2} \sum_{i} 
s_i^z,\nonumber\\
\hspace{-0.4cm}
[d^2 w_i]&=&\frac{1}{\pi}d^3\vec{s}_i\ \delta(\vec{s}_i \cdot \vec{s}_i-1).
\label{O3rep}
\end{eqnarray}

\subsection{Relation between the CP$^1$ model and the other models}
Our model is an extended version of the FXYM, 
\begin{equation}
H_{\rm XY} = - J \sum_{\langle i,j \rangle} \cos (\theta_i - \theta_j + A_{ij}). 
\label{XYM}
\end{equation}
To see this, we rewrite the CP$^{1}$ variables as 
\begin{eqnarray}
w_{i} = 
  \left( 
              \begin{array}{c}
              w_{1i} \\
              w_{2i} \\
              \end{array}
       \right) =  
       \left( 
              \begin{array}{c}
              \cos \left( \psi_{i}/2 \right) e^{i \lambda_{1i}}\\
              \sin \left( \psi_{i}/2 \right) e^{i \lambda_{2i}} \\
              \end{array}
       \right) . \label{spinanglecp}
\end{eqnarray}
Here, the angle variables have the ranges $0 \leq \psi_{i} \leq \pi$, 
$0 \leq \lambda_{1 i,2 i} \leq 2\pi$. The Hamiltonian Eq. (\ref{2denergy}) can be written as 
\begin{eqnarray}
H_{\rm CP^1} = - \frac{t}{4} \sum_{\langle i,j \rangle} \sin \psi_i \sin \psi_{j} 
\cos (\theta_{i} - \theta_{j} + A_{ij}) \nonumber \\
-\frac{\mu}{2} \sum_i \cos \psi_i,  \label{CP1energy} 
\end{eqnarray}
with $\theta_{i} = \lambda_{2i} - \lambda_{1i}$ and 
the integration measure $[d^2 \omega_i] =
(4\pi^2)^{-1} \sin \psi_i d\psi_i 
d\lambda_{1i} d\lambda_{2i}$.
The O(3) spin of Eq. (\ref{O3rep}) is expressed as
\begin{eqnarray}
s_i^x=\sin\psi_i \cos\theta_i,\ s_i^y=\sin\psi_i \sin\theta_i,\ 
s_i^z=\cos\psi_i.
\label{O3psitheta}
\end{eqnarray}
If we restrict the configuration space with fixed $\psi_i=\pi/2$, the ground state has 
(uniform) density $\rho=1/2$, so that $s_i^z =0$ and all 
the spins lie in the $xy$-plane. 
Then, the Hamiltonian reduces to the FXYM [Eq. (\ref{XYM})].
Here, ``frustration'' refers to the fact that,
with $f \neq 0$ for any plaquette, the angles $\theta_i$ around
this plaquette cannot be chosen to maximally satisfy the XY exchange couplings. 
The CP$^1$ model has a site-dependent factor $\sin \psi_i \sin \psi_{j}$ associated with the 
variation of the particle number at each site. Hence, the CP$^1$ model 
includes the particle number fluctuation and goes beyond the 
XY model based on the phase fluctuation only. 

If we take into account the nearest-neighbor repulsive interaction in the 
Bose-Hubbard model, the hard-core constraint yields the gauged 
spin-half quantum XXZ model \cite{Lindner}
\begin{eqnarray}
\hat{H}_{\rm XXZ} =  -\frac{t}{2} \sum_{\langle i,j\rangle} \left( \hat{s}_i^+\hat{s}^-_j e^{i A_{ij}} +
 {\rm c.c,} \right) 
 + V  \sum_{\langle i,j\rangle} \hat{s}_i^z\hat{s}^z_j  \nonumber \\ 
  - \mu\sum_i \hat{s}^z_i.
\end{eqnarray}
Our model corresponds to the classical version of the gauged XXZ model with $V=0$, 
called the XX0 (three-component XY) model \cite{Cuccoli}. 
The XX0 model is clearly distinct from 
the XY (two-component XY) model, as the spins fluctuate also out of the $xy$ plane. 
In other words, even if the XX0 model and the XY model share the same form of 
the Hamiltonian, the associated phase space is different. 

In addition, we note that there is a similar model in which the spins 
are random in not only their direction but also their magnitudes, 
known as the ``fuzzy" spin XY model, \cite{Kawasaki} 
\begin{equation}
H_{\rm FXY} = - J \sum_{\langle i,j \rangle} x_i x_j 
\cos (\theta_{i} - \theta_{j}). 
\end{equation}
%with  $0 \leq \psi_{i} \leq \pi$. Although the form of this model is similar to 
%the CP$^1$ model (\ref{CP1energy}), its property is again very different. 
%Here, $x_i$ gives the magnitude of spin.
While the magnitudes $x_i$ of spins in the fuzzy XY model can have any values 
randomly and continuously site by site, the magnitude 
of O(3) spin $\vec{s}_{i}$ at each site 
in the CP$^1$ model is fixed  unity, as shown in Eq. (\ref{O3rep}). 

In the following, we compare the thermodynamic properties of our gauged CP$^1$ 
model Eq. (\ref{CP1energy}) and the FXYM Eq. (\ref{XYM}). The latter has
been  extensively studied for decades 
\cite{TJ83,LKG,Santiago-Jose,LeeLee,Olsson,Luo,Korshunov,HPV05a,Olsso1n-Teitel,Minnhagen,Okumura}. 
To this end, we have to put the two models in the same energy measure
by establishing a relation between $J$ and $t$. 
By noting the argument given below Eq. (\ref{O3psitheta}), 
let us try to replace the density at site $i$  in Eq. (\ref{CP1energy})
to the average value $\rho$, i.e., 
$\sin \psi_i \sin \psi_{j} \to \sin^2 \psi$ with
\begin{equation}
 \sin^2 \psi = 4 \cos^2\frac{\psi}{2} \left( 1-\cos^2\frac{\psi}{2} \right) 
=4 \rho (1-\rho),
\end{equation}
where we have used $\rho \simeq \langle \phi^{\ast} \phi \rangle 
=  \langle w_{1}^{\ast} w_{1} \rangle =  \cos^{2} (\psi/2) $. 
This correspondence implies the relation 
$J=t \rho(1-\rho)$. We use this relation when the energies 
in the two models (\ref{XYM}) and (\ref{CP1energy}) are compared. 
Note that this relation reflects the particle-hole symmetry and
is {\it different from} the naive replacement $J=t \rho$.

\section{Ground state}\label{ground}
We discuss here the ground-state properties 
of the CP$^1$ model. It is expected that the 
ground state exhibits similar behaviors to the FXYM, where the 
magnetic flux forms the typical pattern with the $q \times q$ unit cell 
structures for $f=p/q$; the checkerboard 
pattern emerges for $f=1/2$ and the staircase pattern 
for $1/3 \leq f \leq 1/2$ \cite{Teitel,Halsey1,Straley}. 
Since the CP$^1$ model has an additional degree of freedom associated with the 
density fluctuation, it is expected that there are some differences from 
the results of the FXYM. 

\begin{figure}
\centering
\includegraphics[width=1.0\linewidth]{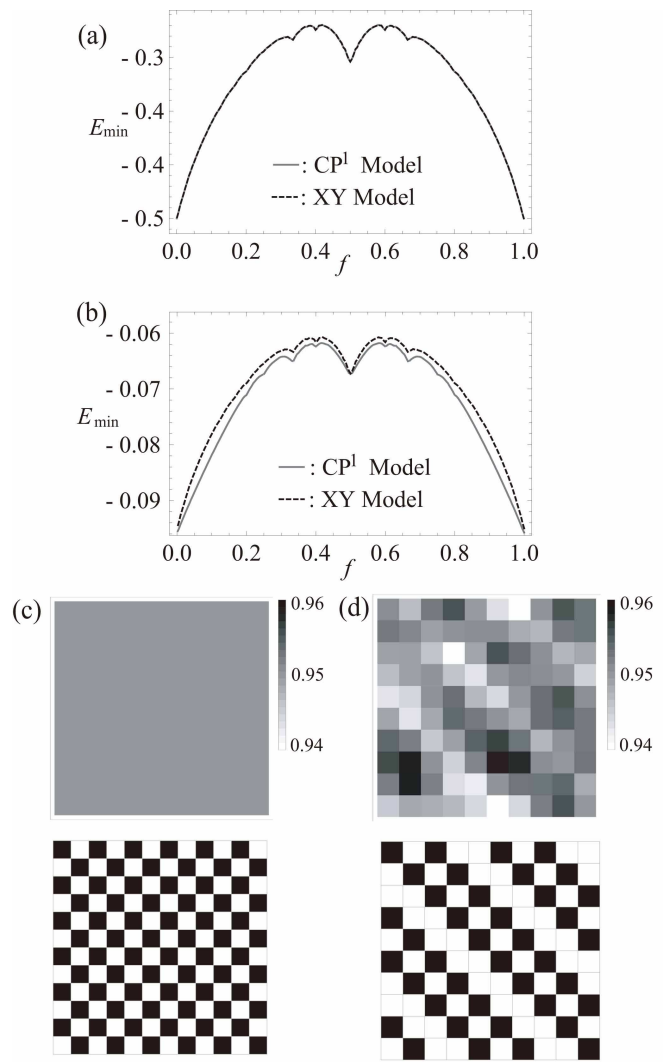}
\caption{The ground-state energy per site 
of the CP$^1$ model Eq. (\ref{CP1energy}) 
as a function of the magnetic flux $f$ for the averaged density (a) $\rho =0.5$ and (b) 0.95, 
obtained through the simulated annealing. 
The solid curve denotes the energy of the CP$^1$ model without the chemical-potential term, 
while the dashed curve denotes that of the XY model 
with setting $J = t\rho(1-\rho)$ for comparison. The two curves 
overlap completely in (a); see the text for the reason. 
(c) and (d) show the distribution over the lattice ($L=12$ and 10 for $f=1/2$ and 2/5, respectively) 
of the mean density 
$\langle \rho_{\bar{i}} \rangle$ (the upper panels)
and the vorticity $m_{\bar{i}}$  (the lower panels) at each plaquette $\bar{i}$ (the 
site of the dual lattice) for $\rho=0.95$; 
(c) $f=1/2$ and (d) $f=2/5$.}
\label{groundenerg} 
\end{figure}
We calculate numerically the ground state  
of the gauged CP$^1$ model 
for fixed $\rho$ by the simulated annealing method. 
Two examples for $\rho=0.5$ and 0.95 are shown in Fig. \ref{groundenerg}, 
where 
the former corresponds to $\mu=0$ (see Appendix \ref{app1}) while the latter is obtained 
by adjusting a proper value of $\mu$ for a given $f$. 
Figures \ref{groundenerg} (a) and \ref{groundenerg} (b) represent the ground-state 
energy $E_{\rm min}$ as a function of $f$. 
We also plot the energy for the FXYM for comparison. 
The shape of the energy curve is non-monotonic behavior with respect to $f$, following 
the bottom of the energy spectrum of the Hofstadter butterfly \cite{Straley}. 

For $\rho =0.5$ the ground-state energy coincides completely with that 
of the XY model. 
This is clear because  $\psi_i$ is freezed to $\pi/2$ at $\rho =0.5$
and the particle-number does not fluctuate. Because $s_i^z=0$ there, 
the ground-state properties are not affected at all even for $f \neq 0$.
As $\rho$ deviates from $0.5$, the ground-state energy of 
the CP$^1$ model becomes slightly lower than that of the XY model, 
except for $f=0.5$, as shown in Fig. \ref{groundenerg}(b). 

Figures \ref{groundenerg}(c) and \ref{groundenerg}(d) show the distribution of the mean density $\rho_{\bar{i}}$ and 
the vorticity $m_{\bar{i}}$ at the site of the dual lattice $\bar{i}$, defined by  
\begin{eqnarray}
\rho_{\bar{i}} = \frac{1}{4} \sum_{i \in \alpha} \rho_{i}\quad
m_{\bar{i}} =\frac{1}{2 \pi} \sum_{i,j \in \alpha} ( \theta_{i} - \theta_{j} + A_{ij}),
\end{eqnarray}
where $|\theta_{i} - \theta_{j} + A_{ij}| \leq \pi$. 
The pattern of $m_{\alpha}$ constitutes the structure of a $q \times q$ unit cell, being 
similar to that of the FXYM. 
As discussed above, the ground-state particle density for $\rho = 0.5$
becomes uniform for any values of $f$. 
For $\rho \neq 0.5$, however, the mean density is also modulated spatially 
in accordance with the distribution of vorticity except for $f=1/2$. 
This is the reason why the ground-state energy of the CP$^1$ model is lower 
than the XY model. 

\section{Phase structures at finite $T$} \label{phasestructure}
We next turn to the discussion on the finite-temperature phase structures 
of the gauged CP$^1$ model and compare the result with the FXYM of Eq. (\ref{XYM}). 
As described in Sec. \ref{intro}, the FFXYM with $f=1/2$ 
may give rise to a nontrivial double phase transition 
\cite{TJ83,LKG,Santiago-Jose,LeeLee,Olsson,Luo,Korshunov,HPV05a,Olsso1n-Teitel,Minnhagen,Okumura}
associated with the Ising transition apart from the BKT transition 
which is normally present at $f=0$. In addition, there have been some discussions 
on the phase transitions for $f=2/5$ and $f=1/3$ \cite{Grest,Li,Lee,Denniston}; 
their nature is dominated by the properties of the domain walls, following 
the Ising-like transition for $f=1/3$ and the first-order transition for $f=2/5$. 
Our primary interest is to see how an additional degree of freedom, associated with 
the particle number fluctuation at each site, modifies these properties. 
We made multicanonical Monte Carlo simulations \cite{Berg} to calculate 
statistical averages of some quantities described below. 

We study the thermodynamic properties by calculating the specific heat 
defined by
\begin{equation}
%U = \frac{1}{L^2} \langle H_{\rm CP^{1}} \rangle, \hspace{3mm}  
C =  \frac{1}{L^2} \left( \langle H_{\rm CP^{1}}^2 \rangle -  \langle H_{\rm CP^{1}} \rangle^2 \right). 
\end{equation}
%Because the partition function $Z$ of Eq. (\ref{2d}) involves integrations over 
%bosonic variables only, one may apply the standard Monte Carlo simulations
%to evaluate $U$ and $C$. 
This value can be useful to locate the first- and second-order
phase transition. 
In addition, to study the BKT transition, we calculate the in-plane susceptibility 
defined as $\chi = \partial \langle \bar{\phi} \rangle / \partial (\beta h) | _{h \to 0} $ with 
the site-average of the hard-core boson field $\bar{\phi} =  \sum_i \phi_i / V$, explicitly written as 
\begin{eqnarray}
\chi = \frac{1}{L^2} \left\langle  \sum_{i,j} \phi_{i}^{\ast} \phi_{j} \right\rangle 
-  \frac{1}{L^2} \left\langle \sum_{i} \phi_{i}^{\ast} \right\rangle \left\langle \sum_{i} \phi_{i} \right\rangle. 
\label{suscepti}
\end{eqnarray}
Here, an auxiliary term $-h \sum_{i} \phi_i$ was introduced in Eq. (\ref{2denergy}) to derive Eq. (\ref{suscepti}).
The second term of the right-hand side of Eq. (\ref{suscepti}) vanishes because of the global U(1) symmetry.
This value grows as $\chi \propto L^{2-\eta}$ with the system size $L$ when 
the correlation function obeys the power-law behavior $\langle \phi^{\ast}_{i} \phi_{j} \rangle \propto r^{-\eta}$ 
(with $\eta \leq 2$) due to the presence of the quasi-long range order. 
On the other hand, it remains finite for $L \to \infty$ when the correlation decays exponentially as 
$\langle \phi^{\ast}_{i} \phi_{j} \rangle \propto e^{-mr}$. 
The value $\eta$ is an exponent for the in-plane correlations below the BKT 
transition temperature $T_{\rm BKT}$, 
being dependent on the temperature. 
We can calculate the exponent, as a function of temperature, by fitting 
the susceptibility for several system sizes $L$ to the above expression for 
each temperature. For the conventional XY model the critical temperature $T_{\rm BKT}$ can 
be estimated when $\chi$ grows as $L^{7/4}$, i.e., $\eta = 1/4$
%the result of the Monte Carlo simulations gives $T_{\rm BKT} \simeq 0.898J/k_{\rm B}$ \cite{Gupta}. 
%In the FFXYM, the nature of the transition into the superfluid phase has been under debate; 
%the recent large-scale numerical simulations have revealed the BKT-type transition with 
%a non-universal jump with $\eta \simeq 0.20$ and $T_{\rm BKT} \simeq 0.442 J/k_{\rm B}$ \cite{Okumura}. 

Furthermore, we study the helicity modulus which is directly connected to the superfluid density. 
The helicity modulus $\Upsilon$ is a measure of the resistance 
to an infinitesimal spin twist $\Delta \theta$ across the system along one coordinate. 
More precisely, it is defined through the change of the total free 
energy $F$ with respect to an infinitesimal twist on the spin 
configuration along, say, the $x$-axis $\theta_{i} - \theta_{i'} \to 
\theta_{i} - \theta_{i'} + \delta \theta$, where $i'$ is the nearest-neighbor 
site of the $i$-site along the $x$-direction and 
$\delta \theta = \Delta \theta/L$. 
One readily finds 
\begin{equation}
\Delta F = \Upsilon (\delta \theta)^2 + {\cal O}((\delta \theta)^4), 
\end{equation}
and $\Upsilon$ can be given as 
\begin{equation}
\Upsilon = -\frac{1}{L^2} \left( \langle H_x \rangle + \beta \langle I_x^2 \rangle  \right),
\end{equation}
where $H_x$ is the $x$-bond part of the Hamiltonian at $\Delta \theta =0$ and $I_x$ 
is the total current in the $x$-direction. For the CP$^1$ model, they are written as 
\begin{eqnarray}
H_x =  - \frac{t}{4} \sum_{\langle i,j \rangle_x} \sin \psi_i \sin \psi_{j} \cos (\theta_{i} - \theta_{j} + A_{ij}),  \nonumber\\
I_x = - \frac{t}{4} \sum_{\langle i,j \rangle_x} \sin \psi_i \sin \psi_{j} \sin (\theta_{i} - \theta_{j} + A_{ij}).
\end{eqnarray}
According to the renormalization-group theory, the helicity modulus for the conventional XY model 
in an infinite system jumps from zero to the finite value $(2/\pi) k_{\rm B}T_{\rm BKT}$ 
at the critical temperature $T=T_{\rm BKT}$ \cite{Nelson}. Therefore, a rough estimate 
of the critical temperature at a finite system could be obtained simply 
by locating the intersection of $\Upsilon$ as a function 
of $T$ and the straight line $\Upsilon = 2 k_{\rm B}T / \pi$. 

In the FFXYM, the jump size has been suggested to be lager than the universal value 
\cite{TJ83,LKG,Santiago-Jose,LeeLee,Okumura}.
As a more accurate method to estimate both $T_{\rm BKT}$ and the jump size 
in the helicity modulus, an useful finite-size-scaling expression at $T=T_{\rm BKT}$ 
is known as \cite{Weber}
\begin{equation}
\Upsilon(L,T=T_{\rm BKT}) = \frac{2}{\pi} k_{\rm B} T^{\ast}_{\rm BKT} \left( 1+ \frac{1}{2} \frac{1}{\ln L + c} \right) 
\label{helicityscaling}
\end{equation}
with $T^{\ast}_{\rm BKT}$ and $c$ being fitting parameters. 
$T^{\ast}_{\rm BKT}$ is related to the jump size, 
being equal to $T_{\rm BKT}$ in the case 
of the usual XY model. 
By making a fit of the numerical data to Eq. (\ref{helicityscaling}) at various temperatures, 
one can estimate the transition temperature $T_{\rm BKT}$ as well as 
the critical exponent from the jump size 
as $\eta = T_{\rm BKT}/4T_{\rm BKT}^{\ast}$ \cite{Okumura,Minnhagen2}. 
%The fitted parameter $T^{\ast}_{\rm BKT}$ at $T_{\rm BKT}$ times $T_{\rm BKT}$ 
%gives the jump size in the helicity modulus $\Upsilon$.

\subsection{Density fluctuation}
\begin{figure}
\centering
\includegraphics[width=1.0\linewidth]{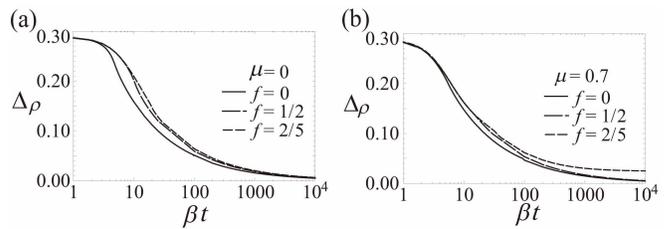}
\caption{The mean density $ \rho $ and the particle number fluctuation 
$\Delta_{\rho}$ as a function of $\beta \mu$ at the high-temperature or zero-hopping limit $\beta t \to 0$; 
the behavior is thus independent of $f$.}
\label{deltaandDeltarho}.  
\end{figure}
Before discussing the detailed thermodynamic properties, let us see 
the magnitude of the spatial fluctuation of the particle density, which 
is the important difference between the gauged CP$^1$ model 
and the FXYM. 
The particle density fluctuation is defined as 
\begin{equation}
\Delta_{\rho} = \sqrt{ \left\langle \frac{1}{L^2} \sum_{i} (\rho_{i} - \bar{\rho})^2 \right\rangle},
\end{equation}
which is zero for the XY model. In the high-temperature limit $\beta t \to 0$, 
one can calculate exactly the partition function 
$Z=[(e^{\beta \mu} -1)/\beta \mu]^V$, and thus
\begin{eqnarray}
\rho = \frac{1+( -1 + \beta \mu) e^{\beta \mu}}{\beta \mu (-1+e^{\beta \mu} )}, \\
\Delta_{\rho} = \frac{-2 + (2 - 2 \beta \mu + \beta^2 \mu^2) e^{\beta \mu}}{\beta^2 \mu^2 (-1+e^{\beta \mu} ) },
\end{eqnarray}
which are shown in Fig. \ref{deltaandDeltarho}. 
As one can see below [Fig. \ref{f=05mu=07}(a)], the mean density $\rho$ 
is weakly dependent on $\beta t$ for $\beta \mu \neq 0$. 
The density fluctuation $\Delta_{\rho}$ has a maximum at $\beta \mu = 0$, 
corresponding to half-filling $\rho =1/2$, 
and decreases for $\beta \mu \to \pm \infty$. Since this $\Delta_{\rho}$ gives an upper limit of the 
expected particle number fluctuation, one can see that influence of the particle 
number fluctuation is less than 20\% in a temperature range of our interest. 

\begin{figure}
\centering
\includegraphics[width=1.1\linewidth]{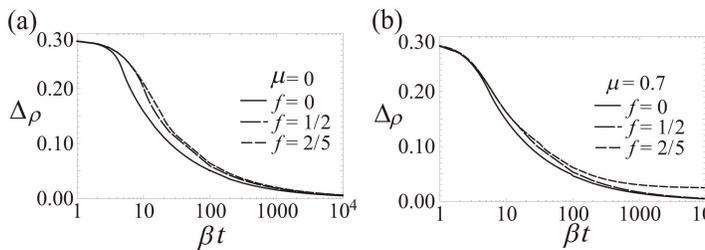}
\caption{The particle number fluctuation $\Delta_{\rho}$ 
as a function of $\beta t$ for (a) $\mu=0$ and (b) $\mu=0.7$. 
The solid, dashed, and dotted curves correspond to $f=0$, 1/2, and 2/5, respectively.}
\label{fluclowtemp}.  
\end{figure}
As seen in the ground state (Sec. \ref{ground}), the density 
becomes uniform for $\mu=0$. Thus, $\Delta_{\rho}$ should go to zero as $\beta t \to \infty$ 
for $ \mu=0$. On the other hands, for $ \mu \neq 0$, it must remains finite 
because the ground state possesses spatial density modulation. 
Figure \ref{fluclowtemp} shows the $\beta t$-dependance of $\Delta_{\rho}$ for $\mu = 0$ and 0.7 
for several values of $f$. 
For $\mu=0$, $\Delta_{\rho}$ approaches to zero as $\beta t \to 0$ for any values of $f$. 
For $\mu=0.7$, on the other hand, $\Delta_{\rho}$ approaches to zero as $\beta t \to 0$ only for $f=1/2$, but 
remains finite for the other values of $f$. 
This behavior is consistent with the ground-state property shown in Fig. \ref{groundenerg}.

\subsection{The case of half-filling ($\mu = 0$)}
Here, we consider the case of half-filling $\rho = 0.5$ by setting $\mu = 0$. 
Then, the density distribution is completely uniform in the ground state, as seen 
in the complete overlap of the ground-state energy in Fig. \ref{groundenerg}(a), 
and thus the CP$^1$ model reproduces the ground state of the XY model. 
However, one has to take into account the density fluctuation at finite temperatures. 
This additional degree of freedom makes the finite-temperature phase diagram 
and the nature of the phase transition nontrivial. 
Here we focus on the case $f=0$, 1/2 and 2/5, each of which has been known to give rise 
to quite different phase transitions in the FXYM.

\subsubsection{f=0}
\begin{figure}
\centering
\includegraphics[width=1.0\linewidth]{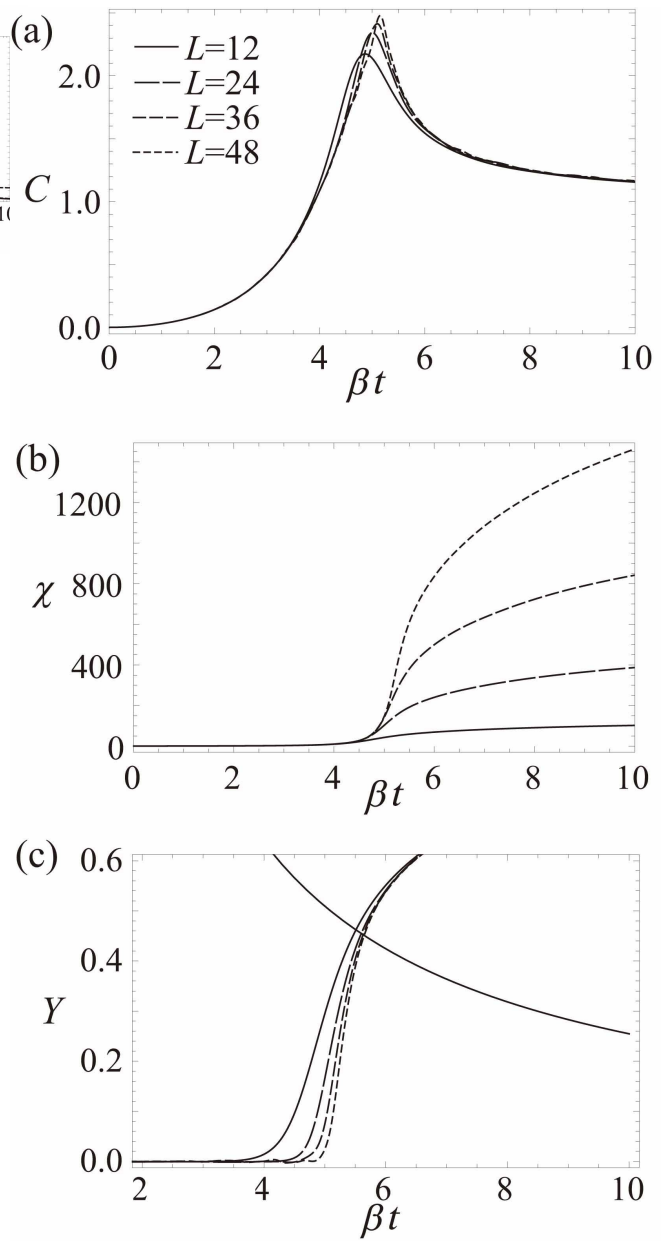}
\caption{Several thermodynamic quantities for $f$=0:  (a) The specific heat $C$, 
(b) the in-plane susceptibility $\chi$, and (c) the helicity 
modulus $\Upsilon$ for the CP$^1$ model with $\mu=0$ and $f=0$ 
as a function of $\beta t $. The system size for each curve is $L=12$, 24, 36, 48. 
An additional curve $\Upsilon = 8/\pi \beta t$ in (c) indicates the universal jump of a BKT transition.}
\label{f=0mu=0}.  
\end{figure}
First, we consider the situation of a zero magnetic field $f=0$, which is useful to 
confirm the accuracy of our numerical computation. 
Then, the model is equivalent to the XX0 (three-component XY) model studied in Ref. \cite{Cuccoli}.
The phase transition of this model has been found to be consistent with the BKT theory. 
The specific heat $C$ has very small finite-size effects. 
The in-plane susceptibility $\chi$ is a strongly increasing function 
of the system size $L$ for low temperatures, while all the data fall on the same curve for high temperatures. 
These facts indicate the absence of second-order transition
and the possibility of the BKT type transition involving the power-law decay
of the correlation function. 
The transition temperature of the XX0 model was obtained as $T_{\rm BKT} = 0.699J/k_{\rm B}$ \cite{Cuccoli}, 
which is lower than the usual (two-component) XY model $T_{\rm BKT} = 0.898J/k_{\rm B}$ \cite{Gupta}.
This is naturally understood due to the difference of the degree of freedom of these models. 

Our numerical result is shown in Fig. \ref{f=0mu=0}. 
The size dependance of $C$ is actually small. 
We estimate the BKT transition temperature $T_{\rm BKT}$ by two methods,
one by using the data of $\chi$ and the other by using $\Upsilon$. 
We plot $\chi/L^{7/4}$ for some values of $L$ as a function of $\beta t$ by assuming $\eta=1/4$, 
in which the crossing point of the curves gives $T_{\rm BKT}$. 
From $\Upsilon$, we take the temperatures that correspond to the crossing 
points of $\Upsilon (T)$ and the line $\Upsilon = (2/\pi) k_{\rm B} T_{\rm BKT}$ related 
to the universal jump value for various values of $L$, interpolating them to $L \to \infty$.
Both these methods give the same $T \approx 0.702 J/k_{\rm B}$, 
which is consistent with the result of Ref. \cite{Cuccoli} 
and confirms the accuracy of our numerical computations.

\subsubsection{f=1/2}\label{f=1/2story}
\begin{figure}
\centering
\includegraphics[width=1.0\linewidth]{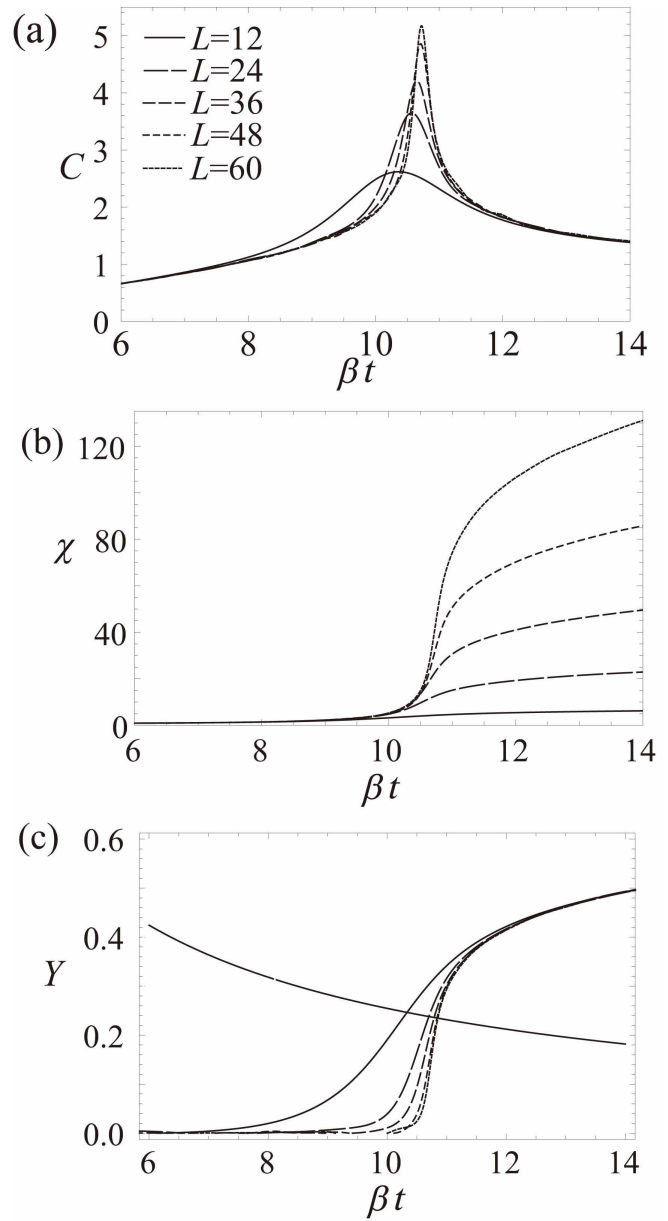}
\caption{Several thermodynamic quantities for $f$=1/2:  (a) The specific heat $C$, 
(b) the in-plane susceptibility $\chi$, and (c) the helicity 
modulus $\Upsilon$ for the CP$^1$ model with $ \mu = 0$ and $f = 1/2$
as a function of $\beta t $. The system size for each curve is $L=12$, 24, 36, 48, 60. 
An additional curve $\Upsilon = 8/\pi \beta t$ in (c) indicates the universal jump of a BKT transition.}
\label{f=05mu=0}. 
\end{figure}
We next consider the case with full frustration $f=1/2$. 
If there is a continuous phase transition, various quantities should exhibit a singular behavior 
near the transition temperature $T_{c}$. 
Figure \ref{f=05mu=0}(a) represents the specific heat $C$
as a function of $\beta t$, where 
$C$ exhibits a peak structure as a function of $\beta t$ and 
the height of the peak increases with increasing $L$. 
This suggests the occurrence of a second-order phase transition. 
Concurrently, $\chi$ and $\Upsilon$ grow from zero with increasing $\beta t$, 
which is a signature of the emergence of superfluid order. 
Hence, the qualitative feature of the phase transition is similar to the FFXYM.

It is important to clarify whether the nature of the phase transition is consistent 
with those observed in the analysis of the FFXYM. 
In the FFXYM, two separate phase transitions may occur, corresponding to 
the breaking of the Z$_2$ chirality and the U(1) symmetry 
\cite{TJ83,LKG,Santiago-Jose,LeeLee,Olsson,Luo,Korshunov,HPV05a,Olsso1n-Teitel,Minnhagen,Okumura}. 
It is still inconclusive whether the former obeys the universality class of the Ising transition 
and the latter is subject to the BKT mechanism with non-universal jump of the helicity modulus. 

First, we focus on the data of the specific heat to clarify the Z$_2$-related phase transition. 
To determine the critical temperature and the critical exponents, 
we use finite-size scaling analysis \cite{FSS}, 
%which extracts values for the critical exponents by 
%investigating how measurements depend on the size $L$ of the system.
%This procedure is carried out by expressing a quantity of interest
%in terms of the correlation length and then introducing a new
%dimensionless function, known as a scaling function.  
where the specific heat $C$ is expressed as
\begin{eqnarray}
C (L,\tilde{t})  =  L^{\alpha/\nu} \tilde{\phi}(L^{1/\nu}\tilde{t}),
\label{eq:gammasf}
\end{eqnarray}
where $\tilde{t}=(T-T_{c})/T_{c}$ is a reduced temperature, $\alpha$ the standard exponent 
for the specific heat, and $\nu$ the exponent for the divergence of the correlation length. 
Since the scaling functions $\tilde{\phi}$ should depend on a single variable, we can make 
all the data for each system size $L$ fall on the same curve by appropriately 
adjusting the values of the critical exponents $\alpha$, $\nu$ and $T_c$. 
For a finite lattice the peak in the specific heat scales with 
system size as $C_{\rm max} \propto L^{\alpha/\nu}$ and occurs at the 
temperature where the scaling function $\tilde{\phi}(L^{1/\nu} \tilde{t})$ is maximum, 
defining the finite-lattice transition temperature $T_c(L)=T_c+{\rm const} \times L^{-1/\nu}$. 

\begin{figure}
\centering
\includegraphics[width=1.0\linewidth]{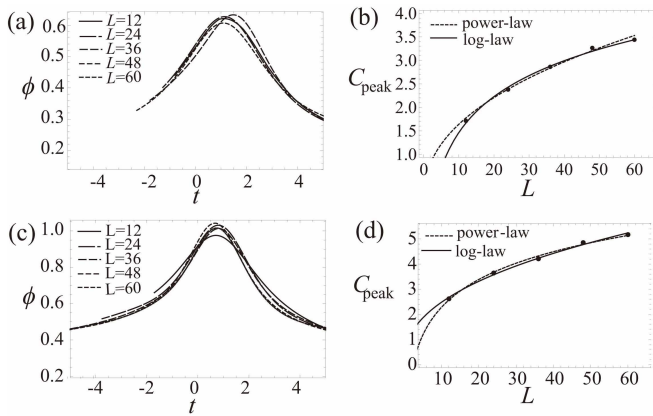}
\caption{The left panels show the power-law scaling collapse of the specific heat data 
for (a) FFXYM and (c) gauged CP$^1$ model.
The right panels show the power-law and logarithmic fitting of the specific heat peak 
with respect to sizes $L$ for (b) FFXYM and (d) the gauged CP$^1$ model. 
The power-law fitted value is $\alpha/\nu = 0.439 $ for (a) and $\alpha/\nu = 0.397 $ for (c).}
\label{f=1/2mu=0heatscaling}. 
\end{figure}
The obtained scaling functions are plotted in Fig. \ref{f=1/2mu=0heatscaling} for both the 
FFXYM and the gauged CP$^1$ model. For the FFXYM, we obtain $T_{c} = 0.454 J/k_{\rm B} $, 
$\nu=0.873$ and $\alpha=0.383$ from Fig. \ref{f=1/2mu=0heatscaling}(a), 
which is consistent with the hyperscaling relation $d \nu = 2-\alpha$ ($d$ is the dimension number). 
Our scaling analyses support the non-Ising exponent $\nu < 1$, 
which is consistent with some of the literature \cite{LKG,Santiago-Jose,LeeLee,Luo}. 
However, it should be noted that there have been several claims that this non-Ising exponent 
is caused by the artifact of the finite-size effect and the exponent in the infinite system may 
be the Ising one $\nu=1$ \cite{Olsson,Olsso1n-Teitel,Okumura}. 
Since our calculation does not have enough system size to resolve this problem and we 
cannot distinguish the data of Fig. \ref{f=1/2mu=0heatscaling}(a) as a power-law fitting or 
logarithmic fitting [see Fig. \ref{f=1/2mu=0heatscaling}(b)], we shall not go to discuss the details of this issue. 
Our main claim in this work is that the similar behavior also occurs for the gauged CP$^1$ model. 
For this model, we also extract the critical exponents from the same analysis for Fig. \ref{f=1/2mu=0heatscaling} 
(c) and (d) as $T_{c} = 0.369 J/k_{\rm B}$, $\nu = 0.878$ and $\alpha = 0.349$. 
The transition temperature becomes lower than that for the FFXYM, while 
the critical exponents are similar. 

\begin{figure}
\centering
\includegraphics[width=1.0\linewidth]{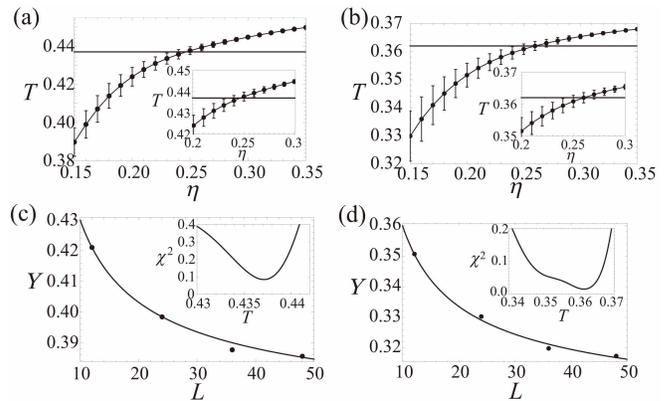}
\caption{Data of the BKT transition of the FFXYM [(a) and (c)] 
and the gauged CP$^1$ model for $f=1/2$ [(b) and (d)]. 
In (a) and (b), we show $T_{\rm BKT}$ obtained by method (i) (see the text)
by the horizontal line; (a) $T=0.437J/k_{\rm B}$ and (b) $T=0.363 J/k_{\rm B}$. 
In addition, the temperature corresponding to the crossing point 
in the method (ii) is plotted as a function of $\eta$. 
Figures (c) and (d) show the best fit of the scaling relation Eq. (\ref{helicityscaling}) 
at the corresponding $T_{\rm BKT}$, which is obtained by finding the 
minimum of $\chi^2$-fit error shown in the inset. 
%(a) $\chi/L^{2-\eta}$ as a function of $T$ for data of Fig. \ref{f=05mu=0}(b) and $\eta=1/4$.
}
\label{f=05mu=0Upsilonscale}. 
\end{figure}
Next, we study the U(1)-related phase transition by employing the same analysis for the $f=0$ case. 
Some studies revealed that the jump size of $\Upsilon$ at $T=T_{\rm BKT}$ 
may be non-universal for the FFXYM \cite{TJ83,LKG,Santiago-Jose,LeeLee,Okumura}. 
Here, we do not assume $\eta = 1/4$ 
%so that the jump size is put as $T_{\rm BKT} / 2\pi \eta(T_{\rm BKT})$ \cite{Okumura}. 
and evaluate $T_{\rm BKT}$ with two different methods: 
(i) Using the $\chi^2$-fit of the scaling relation Eq. (\ref{helicityscaling}), we 
evaluate $T_{\rm BKT}$ and the jump size $T_{\rm BKT}^{\ast}$ at $T_{\rm BKT}$, 
which gives the exponent $\eta = T_{\rm BKT}/4T_{\rm BKT}^{\ast}$. 
(ii) We plot $\chi/L^{2-\eta}$ as a function of $\beta t$ for several system sizes 
$L$ and take the temperature at the crossing point. We make this analysis 
by varying $\eta$ around 1/4 and search the value of $\eta$ that gives 
the same $T_{\rm BKT}$ obtained in the analysis (i). 
%(i) We extract the temperature that corresponds to the crossing 
%point of $\Upsilon$ and the line related to the universal jump value $ T_{\rm BKT} /2 \pi \eta$ 
%for various values of $L$, interpolating them to $L \to \infty$. 
The summary of this analysis 
is shown in Fig. \ref{f=05mu=0Upsilonscale}. For the FFXYM, 
the scaling analysis (i) alone gives $T_{\rm BKT} = 0.437J/k_{\rm B}$ 
and $\eta = 0.2$. This is consistent with the previous literature, 
where $T_{\rm BKT} = 0.437 J/k_{\rm B}$ is slightly lower than $T_{c}$ and 
the BKT jump is non-universal \cite{TJ83,LKG,Santiago-Jose,LeeLee,Okumura}. 
However, the crossing point 
obtained by the analysis (ii) is preferable to $\eta \approx 0.25$, as shown in Fig. \ref{f=05mu=0Upsilonscale}(a),
which suggests the same universality of the conventional BKT transition. 
This usual BKT behavior for the FFXYM was also suggested by Olsson \cite{Olsson}. 
Similar behavior is also found for the gauged CP$^1$ model as 
$T_{\rm BKT} = 0.363 J/k_{\rm B}$ and $\eta \approx 0.22$ for the analysis 
(i) and $\eta \approx 0.26$ for the analysis (ii). 

%Our results  indicate unambiguously that the spin and the chirality exhibit separate phase  transitions 
%at two distinct temperatures, {\it i.e.\/}, the occurrence of the spin-chirality decoupling. 
%The chirality exhibits a long-range order at $T_{\rm c}=0.45324(1)J/k_{\rm B}$ via a second-order phase transition, 
%where the spin remains disordered with a finite correlation  length $\xi_{\rm s}(T_{\rm c}) \sim 120$. 
%The critical properties of the chiral transition determined from a finite-size scaling analysis for large enough 
%systems of linear size $L > \xi_{\rm s}(T_{\rm c})$ are well compatible with the Ising universality. 
%On the other hand, the spin exhibits  a phase transition at a lower temperature $T_{\rm s}=0.4418(5)J/k_{\rm B}$ 
%into the quasi-long-range-ordered phase. We found $\eta(T_{\rm s})=0.201(1)J/k_{\rm B}$, suggesting that the universality 
%of the spin transition is different from that of the conventional Kosterlitz-Thouless (KT) transition.

\subsubsection{f=2/5}
\begin{figure}
\centering
\includegraphics[width=1.0\linewidth]{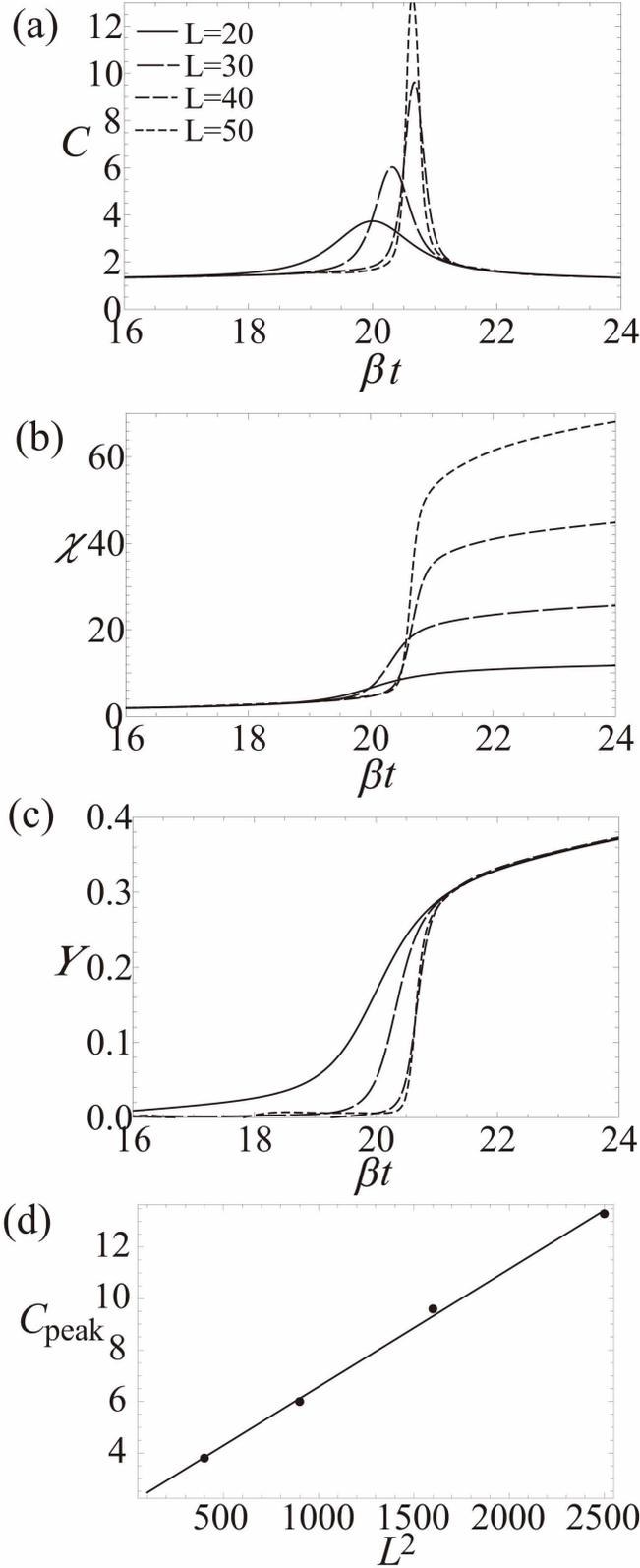}
\caption{Several thermodynamic quantities for the CP$^1$ model with 
$\mu = 0$ and $f = 2/5$ as a function of $\beta t$.
(a) The specific heat $C$, (b) the in-plane susceptibility, 
and (c) the helicity modulus $\Upsilon$. The system size for each curve is $L=20$, 30, 40, 50. 
(d) represents specific heat vs $L^2$}
\label{f=04mu=0scale}. 
\end{figure}
The thermodynamic properties for $f=2/5$ would appear to be similar to 
the $f=1/2$ situation as seen in Fig. \ref{f=04mu=0scale}. 
However, the nature of the phase transition is very different. 
In the FXYM, several works indicated that the transition is associated 
with first-order type \cite{Li,Denniston}. 
Li and Teitel observed hysteresis of the internal energy when the temperature 
was cycled around the transition and used this as an argument for a first-order transition \cite{Li}. 
Denniston and Tang pointed out that the complicated branching structure of domain walls 
is similar to the $q>5$ Pott's models where the first-order transition occurs \cite{Denniston}.
The most direct indication of a first-order transition is the presence of a free energy barrier between 
the ordered and disordered states which diverges as the system size increases \cite{Denniston}. 
Since there is no diverging characteristic length to which the linear dimension $L$ could be compared 
at a first-order transition, one finds that it is simply the volume $L^2$ that controls 
the size effects. 

In the gauged CP$^1$ model, Fig. \ref{f=04mu=0scale}(a) clearly shows the 
rapid growth of the peak of $C$. The inset of Fig. \ref{f=04mu=0scale}(b) 
shows the peak values of $C$ as a function of $L^2$. 
The linear fit clearly shows the expected first-order scaling behavior. 
From the positions of the peaks as a function of $L$, we obtain $T_c =0.193 J/k_{\rm B}$, 
which is again slightly lower than that of the FXYM $T_{c} = 0.2127 J/k_{\rm B}$ \cite{Denniston}. 
The growing $\chi$ and $\Upsilon$ at low temperature certainly 
provides the emergence of the superfluid order. 
Due to the presence of the first-order transition, it is difficult to explicitly discuss 
the properties of the BKT transition. 

%\begin{figure}
%\centering
%\includegraphics[width=1.0\linewidth]{fig6.eps}
%\caption{(color online) The mean density, the specific heat and the in-plane susceptibility for the CP$^1$ 
%model as a function of temperature for $f=0.5$ and $\mu = 10$.}
%\label{f=05mu=10scale}. 
%\end{figure}

\subsection{The case of  non half-filling ($\mu \neq 0$)}
Finally, we show the similar data with the previous subsection but for $\mu \neq 0$, where 
the particle occupation at each site is not half-filling. In terms of the pseudospin 
Hamiltonian (\ref{eq:spham}), this situation corresponds to applying a longitudinal 
magnetic field, as seen in the second term of Eq. (\ref{eq:spham}). 
%For large $\mu$, all spins align along the longitudinal axis, corresponding to 
%the full occupation or full empty of bosons at each site. 
%For intermediate values of $\mu$, although the spins tend to align the longitudinal direction, 
%their reduced transverse component still alive. 

\begin{figure}
\centering
\includegraphics[width=1.0\linewidth]{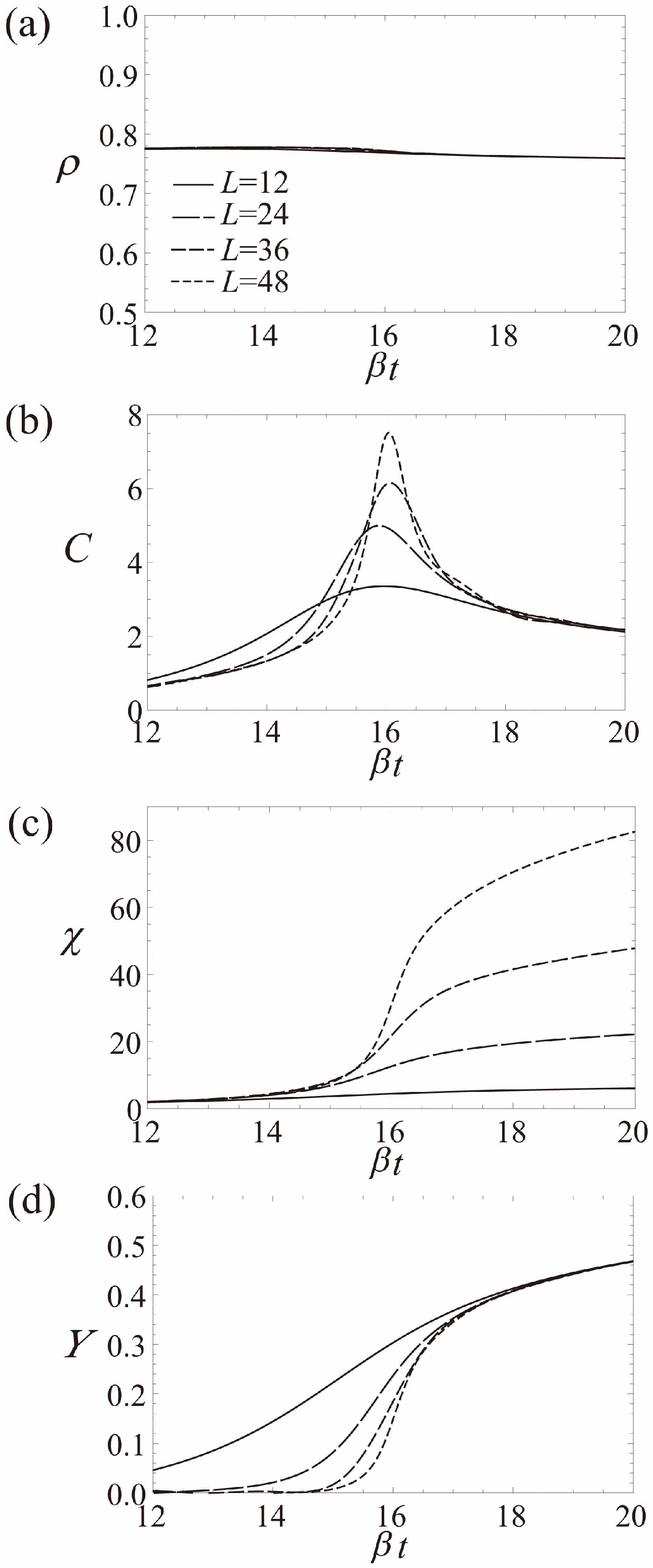}
\caption{Several thermodynamic quantities for the CP$^1$ model with 
$\mu = 0.7$ and $f = 1/2$ as a function of $\beta t$. 
(a) mean particle number $\rho$ 
(b) The specific heat $C$, (c) the in-plane susceptibility $\chi$, 
and (d) the helicity modulus $\Upsilon$. 
The system size for each curve is $L=12$, 24, 36, 48.}
\label{f=05mu=07}. 
\end{figure}
We make the similar analysis as in the previous calculation for $\mu=0.7$, where 
the averaged density is about $\rho \approx 0.7$-$0.8$, which depends 
weakly on $f$ and $\beta t$. The thermodynamic quantities behave similarly to those found in the 
half-filling case; an example for $\mu=0.7$ and $f=1/2$ is shown in Fig. \ref{f=05mu=07}. 
We find no qualitative difference in the phase transition for each $f$ from 
the $\mu=0$ case. 
%As a practical problem, because the amplitude of the transverse spin becomes small, 
%the convergence of the Monte Carlo simulations becomes worse. 
We show in Table \ref{tab:table1} the obtained critical exponents $\nu$, $\alpha$ and $\eta(T_{\rm BKT})$ 
and the critical temperatures $T_{c}$ and $T_{\rm BKT}$ associated with the Z$_2$ 
and U(1) symmetry breaking, respectively. One can see that 
the transition temperature is further reduced from the $\mu=0$ case. 
Also, the critical exponents are modified from the values of $\mu=0$.
This is naturally understood as follows.
As one increases $|\mu|$, the average magnitude of the XY spin component 
$[(s_i^x)^2+(s_i^y)^2]^{1/2}$ decreases because  $|s_i^z|$ increases
(For example, the limit $\mu\to\pm\infty$ implies $s_z=\pm 1$).
If the length of XY spin is fixed, the model should be in the same 
universality class as the FXYM. However, our study for $\mu=0$ and $f=1/2$ 
exhibits that the fluctuations of $s_{i}^{x,y}$ give rise to critical 
exponents different from those of the FFXYM. 
Therefore fluctuations around the XY spins of
different length may certainly produce different critical exponents.  
\begin{table*}
\caption{\label{tab:table1} List of the transition temperatures and some critical exponents obtained in this work. 
The transition temperature is measured by using $J = t \rho (1-\rho)$; for $\mu \neq 0$, $\rho$ is used 
at the corresponding transition temperature. 
In $\eta(T_{\rm BKT})$, we represent two values obtained by the method (i) and (ii) in 
Sec. \ref{f=1/2story}.
For comparison, the corresponding values for the 2D Ising model are $\nu=1$ and $\alpha=0$, which implies 
the logarithmic divergence of the specific heat.}
\begin{ruledtabular}
\begin{tabular}{cccccc}
 & $T_{\rm c}$ & $ \nu $ & $\alpha$ & $T_{\rm BKT}$ & $\eta(T_{\rm BKT})$  \\
\hline
CP$^1$ model, $f=0$, $\mu=0$ &  --- &  --- &  --- & 0.702(1) $J/k_{\rm B}$ & 0.25 \\
CP$^1$ model, $f=1/2$, $\mu=0$ & 0.369(4) $J/k_{\rm B}$  & 0.878(5) & 0.349(9) & 0.363(3) $J/k_{\rm B}$ & (i) 0.22, (ii) 0.26 \\
CP$^1$ model, $f=2/5$, $\mu=0$ & 0.193(1) $J/k_{\rm B}$ &  --- &--- & --- &--- \\
CP$^1$ model, $f=0$, $\mu=0.7$ &  ---  & --- & ---  & 0.672(1) $J/k_{\rm B}$ & 0.25 \\
CP$^1$ model, $f=1/2$, $\mu=0.7$ & 0.347(1) $J/k_{\rm B}$ & 0.781(25) & 0.381(7) & 0.334(4)$J/k_{\rm B}$ & (i) 0.19, (ii) 0.24 \\
CP$^1$ model, $f=2/5$, $\mu=0.7$ & 0.168(1) $J/k_{\rm B}$ &  --- & --- & --- & --- \\
%2D Ising model & 2.269 $J/k_{\rm B}$  & 0 (log) & 1 & --- & ---  \\
2D XY model & --- & --- & --- & 0.898(1) $J/k_{\rm B}$  & 0.25 \\
2D FFXYM & 0.454(1) $J/k_{\rm B}$ &0.873(3) & 0.383(34)  & 0.437(3) $J/k_{\rm B}$ & (i) 0.2, (ii) 0.25
\end{tabular}
\end{ruledtabular}
\end{table*}

%\begin{figure}
%\centering
%\includegraphics[width=1.0\linewidth]{fig9.eps}
%\caption{Several thermodynamic quantities:  (a) the internal energy $U$, 
%(b) The specific heat $C$, (c) the in-plane susceptibility $\chi$, and (d) the helicity 
%modulus $\Upsilon$ for the CP$^1$ model with $\mu=0.7$ and $f=2/5$ 
%as a function of $\beta t $. The system size for each curve is $L=12$, 24, 36, 48.}
%\label{f=04mu=07scale}. 
%\end{figure}

%\begin{figure}
%\centering
%\includegraphics[width=1.0\linewidth]{fig10.eps}
%\caption{(color online) The mean density, the specific heat and the in-plane susceptibility for the CP$^1$ 
%model as a function of temperature for $f=0.5$ and $\mu = 10$.}
%\label{f=05mu=10scale}. 
%\end{figure}

\section{Conclusion and discussion}\label{conclusion}
We study the finite-temperature phase structures of hard-core bosons in a two-dimensional optical lattice
subject to an effective magnetic field by employing the gauged CP$^1$ model.
Based on the multicanonical Monte Carlo simulations, we study their phase 
structures at finite temperatures for several values of the magnetic flux 
per plaquette of the lattice and mean particle density. 
A summary of this work is listed in Table \ref{tab:table1}. 
Also, the magnitudes of the particle density fluctuations are measured
to be less than 20\%. 
They cause the shift of transition temperatures $T_c$ and $T_{\rm BKT}$,
which are slightly decreased from those of the FXYM, and 
also the shift of critical exponents $\alpha$ and $\nu$.
However these fluctuations do not modify the global phase structure
and the critical properties of FXYM.
 
The regime described by the XY model (Josephson junction array) can be realized 
when the mean particle number at each site is very large $\rho_i  \gg 1 $ 
\cite{Kasamatsu2,Trombettoni}, where the particle number fluctuation becomes 
negligible as $\sim \rho_i^{-1/2}$. The BKT transition in such a regime was observed 
by Schweikhard et al. \cite{Schweikhard}.
The important message of this work is that, even though the strong particle number 
fluctuation becomes remarkable due to the small site occupation $\rho_i  \sim 1 $, 
one can expect similar thermal phase transitions seen in the FXYM. 
The recent experimental demonstration on generating an effective magnetic field 
in an optical lattice \cite{Aidelsburger} opens the door to explore the rich finite-temperature phase diagram 
of this system.  

%Below is  the list of tasks Nakano-kun should complete:
%(They are in the order of appearence in the text and not of importance.)
%  \begin{itemize}
%  \begin{enumerate}
%  \item Calculate the ground state energy of $E_{{\rm 2D}}(f,\delta)$ 
%  \item Calculate the phase diagram of $Z_{{\rm 2D}}$ in the $\delta-T$ plane for each $f$  \item Calculate the phase diagram of $Z_{{\rm 2DXY}}$ along $T$ for each $f$
%  (or in the $f-T$ plane)
%  \item Calculate the phase diagram of $Z_{{\rm 3D}}$ in the $\delta-T$ plane for each $f$  \item Calculate the phase diagram of $Z_{{\rm 3DXY}}$ along $T$ for each $f$   
%  (or in the $f-T$ plane)
% \item (unnecessary?) Calculate  the ground-state energy of $E_{{\rm 3D}}(f,\delta)$ 
% \end{enumerate}

\begin{acknowledgments}
The authors thank Shu Tanaka for useful discussions.
One of the authors (K.K.) is supported in part by a Grant-in-Aid
for Scientific Research (Grant No. 21740267) from MEXT, Japan.
\end{acknowledgments}
%\newpage %Just because of unusual number of tables stacked at end

\begin{thebibliography}{99}
\bibitem{gauge}
We use the term ``gauged" to describe the constant gauge field $A_{ij}$
in the present model. This usage  should be distinguished
from the related ``gauge models" \cite{Nakano, Shimizu} in which a 
dynamical U(1) gauge field is involved.
\bibitem{O3}
We note that definition of the O(3) spin vector $\vec{s}_i$ 
has no factor 1/2 of
the SU(2) operator $\hat{s}_i^{x,y,z}$ of Eq. (\ref{spinwrelation})
to ensure $\vec{s}_i \vec{s}_i=1$.
Even if we consider (1/2)$\vec{s}_i$, its squared magnitude
(1/4)$\vec{s}_i \vec{s}_i$=1/4 is different from 3/4 in 
Eq. (\ref{spinwrelation}) reflecting the quantum fluctuations in
the expression $s(s+1)=s^2+s$.
\end{thebibliography}

\appendix

\section{Symmetry of the gauged CP$^1$ model}\label{app1}
We summarize the symmetry properties of the gauged CP$^1$ model 
Eq. (\ref{2denergy}), from which one can get some useful information 
to understand the results. 

The model equation (\ref{2denergy}) or (\ref{O3rep}) has the following symmetry properties: 
\begin{enumerate}
\item The model has global U(1) symmetry; it is invariant under the change 
$\phi_{i} \rightarrow \phi_i '= e^{i \theta} \phi_i$.

\item The model also has local gauge symmetry. 
If we change the gauge ${\bf A} \rightarrow {\bf A} + \nabla \chi$, 
then the Hamiltonian remains unchanged if the boson picks up a phase change as 
$A_{ij} \rightarrow A_{ij}  + (\chi_j - \chi_i)$ 
and $\phi_{i} \rightarrow e^{i\chi_i} \phi_i$. 
In terms of the pseudospin $\vec{s}_i$ for the $i$-th site, 
this corresponds to a rotation with the angle $\chi_{i}$ 
in the $xy$ plane, $s_i^{\pm}\to e^{\mp i\chi_i} s_i^{\pm}$. 
Because the choice $\chi_i = \pi i_x i_y$ gives rise to 
a shift of fluxes $f \to f +1$, the system is periodic in $f$ with the period 1. 
%Hence we restrict $f$ to the interval $0 \leq f < 1$ below.

\item
The Hamiltonian Eq. (\ref{2denergy}) is invariant under the change 
$f \to f' = -f$, corresponding to the time reversal operation 
$\phi_i  \to \phi' = \phi_i^{\ast}$ ($\lambda_{ai} \to \lambda_{ai}' = - \lambda_{ai}$). 
Since the partition function is not affected by this transformation, 
one can show that the internal energy 
$E=\langle H \rangle = - \partial \ln Z / \partial \beta$ 
and the mean number density 
$ \rho = \langle N \rangle / L^2 =L^{-2} \partial \ln Z / \partial (\beta \mu)$ 
have the following properties 
\begin{eqnarray}
E(\beta, t, \mu, f) =  E(\beta, t, \mu, -f),  \label{appmain1} \\
\rho(\beta,t,\mu,f) = \rho(\beta,t,\mu,-f) . \label{appmain2}
\end{eqnarray}
The symmetric form of the ground state energy shown in Fig. \ref{groundenerg} 
can be understood as follows. The plotted energy $E_{\rm min}$ in Fig. \ref{groundenerg} is 
the first term of Eq. (\ref{2denergy}) $E = E_{\rm min} - \mu \rho L^2$. 
From Eq. (\ref{appmain2}), one can see $\mu(\rho(f),\beta,t,f) = \mu(\rho(-f),\beta,t,-f)$. 
For the fixed $\rho$, Eq. (\ref{appmain1}) gives 
\begin{eqnarray}
E_{\rm min} (t,\mu(f),f) = E_{\rm min} (t,\mu(-f),-f) \nonumber \\ 
= E_{\rm min} (t,\mu(1-f),1-f).
\end{eqnarray}
The last equality is due to the invariance for $f \to f+1$. 

\item
Let us consider the transformation $\psi_i \to \psi_i'=\pi-\psi_i$. 
Then, Eq. (\ref{CP1energy}) yields 
$H_{\rm CP^{1}}(\mu) \to H_{\rm CP^{1}}'(\mu) 
= H_{\rm CP^{1}}(-\mu) $ and $\rho= \langle \cos^2(\psi/2) \rangle \to 
\rho'(\mu)=1 - \rho(-\mu) $ (or equivalently 
$s_i^z \rightarrow s_{i}'^{z} = - s_i^z$). 
This reflects the particle-hole symmetry, 
where $\rho'$ should be interpreted as the hole 
density if $\rho$ represents the particle density. 
%\begin{eqnarray}
%\rho (\beta, t, \mu, f) = 1 -\rho (\beta, t, -\mu, f) , \label{appmain2d}
%\end{eqnarray}
Therefore, especially for $\mu=0$, we obtain the relation 
\begin{equation}
\rho (\beta,t,0,f) = \rho'(\beta,t,0,f)  = 1- \rho(\beta,t,0,f)=  \frac{1}{2}.
\end{equation}

\end{enumerate}

%First, let us see the quantum Hamiltonian Eq. (\ref{hardcorehamil}) and its 
%property for the particle-hole exchange. 
%If one changes as $\hat{\phi}_{i} \to \hat{\phi}_i' = \hat{\phi}_{i}^\dag$, the Hamiltonian becomes 
%\begin{equation}
%H(t,\mu,f,\{ \phi' \}) = H(t,-\mu,-f,\{ \phi \}) - \mu L^2. 
%\label{phhamilch}
%\end{equation}
%Thus, this exchange involves the inversion of a magnetic field ($f \to -f$) and 
%the additional constant term $- \mu L^2$.
%Then, the partition function $Z \to Z'$ can be written as 
%\begin{equation}
%Z'(\beta,t,\mu, f) = e^{\beta \mu L^2} Z(\beta,t,-\mu, -f).
%\end{equation}

\end{document}